\documentclass[11pt,draftcls,onecolumn]{IEEEtran}
\usepackage{epsf}
\usepackage{graphics}
\usepackage{times}
\usepackage{graphicx}
\usepackage{amsmath}
\usepackage{amssymb}
\usepackage{epsfig}
\usepackage{subfigure}
\usepackage{array}
\usepackage{float}
\begin{document}

\title{Adaptive Generation Method of OFDM Signals in SLM Schemes for Low-complexity}

\author{Kee-Hoon Kim, Hyun-Seung Joo, Jong-Seon No, and Dong-Joon Shin
\thanks{K.-H. Kim, H.-S. Joo, and J.-S. No are with the Department of Electrical Engineering and Computer Science, INMC, Seoul
National University, Seoul, 151-744, Korea (email: kkh@ccl.snu.ac.kr, joohs@ccl.snu.ac.kr, jsno@snu.ac.kr).}
\thanks{D.-J. Shin is with the Department of Electronic Engineering, Hanyang University, Seoul, 133-791, Korea (email:
djshin@hanyang.ac.kr).}
\thanks{}
}
\maketitle
\begin{abstract}
There are many selected mapping (SLM) schemes to reduce the peak-to-average power ratio (PAPR) of orthogonal frequency division multiplexing (OFDM) signals. Beginning with the conventional SLM scheme, there have been proposed many low-complexity SLM schemes including Lim's, Wang's, and Baxely's SLM schemes typically.

In this paper, we propose an adaptive generation (AG) method of OFDM signals in SLM schemes. By generating the alternative OFDM signals adaptively, unnecessary computational complexity of SLM schemes can be removed without any degradation of their PAPR reduction performance. In this paper, we apply the AG method to various SLM schemes which are the conventional SLM scheme and its low-complexity versions such as Lim's, Wang's, and Baxely's SLM schemes. Of course, the AG method can be applied to most of existing SLM schemes easily. The numerical results show that the AG method can reduce their computational complexity substantially.
\end{abstract}


\section{Introduction}
Orthogonal frequency division multiplexing (OFDM) is a multicarrier modulation
method utilizing the orthogonality of subcarriers. OFDM has been
adopted as a standard modulation method in several wireless communication systems such as digital audio broadcasting (DAB),
digital video broadcasting (DVB), IEEE 802.11 wireless local area network (WLAN),
and IEEE 802.16 wireless metropolitan area network (WMAN).
Similar to other multicarrier schemes, OFDM has a high peak-to-average
power ratio (PAPR) problem, which makes its straightforward implementation quite costly.
Thus, it is highly desirable to reduce the PAPR of OFDM signals.

Over the last decades, various techniques to reduce the PAPR of OFDM signals have been proposed such as clipping~\cite{Oneal},\cite{Ochiai}, coding~\cite{Tsai}, active constellation extension (ACE)~\cite{Krongold}, tone reservation (TR)~\cite{Tellado}, partial transmit sequence (PTS)~\cite{muller}, and selected mapping (SLM)~\cite{Bauml}.

Among them, SLM and PTS are widely used because they show good PAPR reduction performance without bit error rate (BER) degradation. However, they require many inverse fast Fourier transforms (IFFTs), which cause high computational complexity.

It is well known that SLM scheme is more advantageous than PTS scheme if the amount of side information (SI) is limited. However, the computational complexity of SLM scheme is larger than that of PTS scheme. Therefore, many modified SLM schemes with low-complexity have been proposed \cite{Ghassemi}--\cite{Baxely}. We review the representative low-complexity SLM schemes among them.

Firstly, Lim proposed the low-complexity SLM scheme exploiting the signals at an intermediate stage of IFFT in \cite{Lim}. In \cite{Lim}, the signals at an intermediate stage of IFFT are multiplied by phase rotation vectors designed to do not destroy the orthogonality between subcarriers. Secondly, Wang proposed the low-complexity SLM scheme using conversion matrices in \cite{Wang0} and \cite{Wang}. In \cite{Wang}, one IFFT block is required to generate the original OFDM signal sequence and it is converted to many alternative OFDM signal sequences by multiplying conversion matrices. Thirdly, in \cite{Baxely}, Baxely proposed the low-complexity SLM scheme exploiting the characteristics of high power amplifier (HPA) in OFDM systems. That is, Baxely's SLM scheme only tests phase rotation vectors until an OFDM signal sequence with PAPR less than the saturation point of HPA is found.

In this paper, an adaptive generation (AG) method of alternative OFDM signals in SLM schemes is proposed. This methodology can be applied to almost all existing SLM schemes including the conventional SLM scheme and its low-complexity versions such as Lim's, Wang's, and Baxely's SLM schemes. Aided by the proposed AG method, the SLM schemes can be implemented with lower computational complexity than their computational complexity without the AG method. Numerical results show that the AG method can reduce the computational complexity of the SLM schemes substantially although the AG method is applied to the low-complexity SLM schemes which are already modified to have low-complexity. It is meaningful because the proposed AG method do not degrade any PAPR reduction performance of the SLM schemes.

The rest of this paper is organized as follows. In Section II, the conventional SLM scheme and its representative three low-complexity versions are reviewed. In Section III, we introduce the AG method and apply it to the conventional SLM scheme. And we analyze the computational benefit of the AG method stochastically. In Section IV, we briefly describe the applications for the low-complexity SLM schemes. The computational benefit of the AG method is evaluated through the numerical results in Section V and conclusions are given in Section VI.

\section{SLM Schemes : Conventional SLM Scheme and Low-complexity SLM Schemes}
In this section, we briefly introduce OFDM systems, the conventional SLM scheme, and its representative low-complexity versions such as Lim's, Wang's, and Baxely's SLM schemes. We handle these three low-complexity SLM schemes because these SLM schemes are frequently cited in many literatures and have good PAPR reduction performance with reduced computational complexity.

\subsection{Conventional SLM Scheme}
In this paper, we use the upper case $\mathbf{X}=\{X(0),X(1),...,X(N-1)\}$ for the input symbol sequence and the lower case $\mathbf{x}=\{x(0),x(1),...,x(N-1)\}$ for the OFDM signal sequence, where $N$ is the number of subcarriers. The relation between the input symbol sequence $\mathbf{X}$ in frequency domain and the OFDM signal sequence $\mathbf{x}$ in time domain can be expressed by IFFT as
\begin{equation}
\mathbf{x}=\mathrm{IFFT}(\mathbf{X})
\end{equation}
or
\begin{equation}\label{eq:IFFTeq}
x(n)=\sum_{k=0}^{N-1}X(k)W^{-kn}
\end{equation}
where $W=e^{-j\frac{2\pi}{N}}$ and $0 \leq n \leq N-1$.

\begin{figure}[H]
\centering
\includegraphics[width=0.9\linewidth]{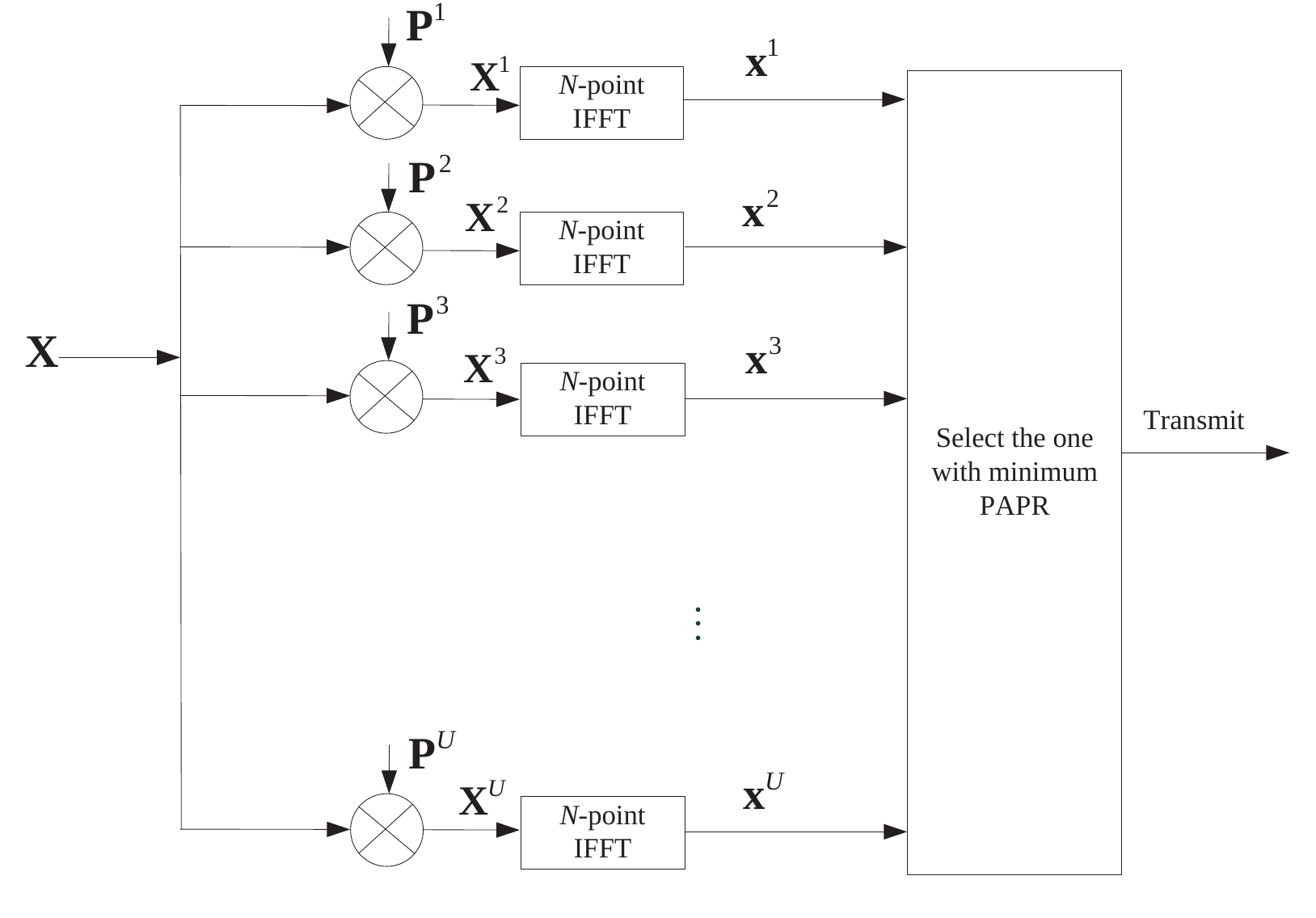}
\caption{A block diagram of the conventional SLM scheme.}
\label{fig:convSLM}
\end{figure}
The conventional SLM scheme in \cite{Bauml} is described in Fig.~\ref{fig:convSLM}, which generates $U$
alternative OFDM signal sequences ${\mathbf x}^{u}$,
$1\le u \le U$. To generate $U$ alternative OFDM signal sequences, $U$ distinct phase rotation vectors ${\mathbf P}^u$ known to both transmitter and receiver are used, where ${\mathbf P}^{u}$=$\{P^u(0),~P^u(1),\cdots,~P^u(N-1)\}$ with $P^u(k)=e^{j\phi^u(k)}$, $\phi^u(k)\in [0,~2\pi )$, $1\le u\le U$. ${\mathbf P}^1$ is the all-one vector for generating the original OFDM signal sequence and thus $\mathbf{x}^1=\mathbf{x}$. The input symbol sequence ${\mathbf X}$ is multiplied by each phase rotation vector ${\mathbf P}^u$ element by element. Then the input symbol sequence ${\mathbf X}$ is represented by $U$ different alternative input symbol sequences ${\mathbf X}^u$, where ${X}^u(k)= { X}(k){ P}^u(k)$, $1\leq u\leq U$.

These $U$ alternative input symbol sequences are IFFTed to generate $U$ alternative OFDM signal sequences~${\mathbf x}^{u}= {\rm IFFT}( {\mathbf X}^u )$ and their components' powers and PAPR values are calculated. Finally, the alternative OFDM signal sequence ${\mathbf x}^{\tilde u}$ having the
minimum PAPR is selected for transmission as
\begin{equation}
\tilde{ u}=\underset{\scriptscriptstyle 1\le u \le U}{\rm arg~min}
\Bigg(\frac{\mathrm{max}| {x}^u(n) | ^2}{E[| {x}(n) | ^2]}\Bigg).
\end{equation}
Note that the SI on $\tilde{ u}$ needs to be transmitted in order to properly demodulate the received OFDM signal sequence at the receiver.

\begin{flushleft}
\rule{.99\linewidth}{0.2mm}\\
\end{flushleft}
\textbf{Pseudo code 1: the conventional SLM scheme}\\
1:~~$\gamma \Leftarrow \infty$.\\
2:~~\textbf{for} $u=1, 2, \cdots, U$\\
3:~~~~~~~generate $\mathbf{x}^u$ processing one $N$-point IFFT.\\
4:~~~~~~~\textbf{if} PAPR of $\mathbf{x}^u < \gamma$\\
5:~~~~~~~~~~~$\gamma \Leftarrow$ PAPR of $\mathbf{x}^u$.\\
6:~~~~~~~~~~~$\mathbf{x}^{\tilde{u}} \Leftarrow$ $\mathbf{x}^u$.\\
7:~~~~~~~\textbf{end if}\\
8:~~\textbf{end for}(u)\\
9:~~transmit $\mathbf{x}^{\tilde{u}}$.\\
\rule{.99\linewidth}{0.2mm}\\

For accurate understanding of the conventional SLM scheme, pseudo code for the conventional SLM scheme is given as Pseudo code 1. The value of $\gamma$ in Pseudo code 1, which is called as \textit{intermediate minimum PAPR value}, is the minimum value among the PAPR values of the alternative OFDM signal sequences already generated.

\subsection{Low-complexity SLM Schemes}

There have been many low-complexity SLM schemes which are modified versions of the conventional SLM scheme. We briefly review the representative three low-complexity SLM schemes. They have lower computational complexity than that of the conventional SLM scheme with the same number of alternative OFDM signal sequences.

\subsubsection{Lim's SLM Scheme \cite{Lim}}
It is already known that one $N$-point IFFT consists of $n=\log_2N$ stages. In Lim's SLM scheme, the $N$-point IFFT is processed from $1$-st stage to $(n-r)$-th stage, not the $n$-th stage, to the input symbol sequence $\mathbf{X}$, where $r$ is the number of remaining stages. And $U$ phase rotation vectors, which are designed to do not destroy the orthogonality between the subcarriers, are multiplied to the output from the $(n-r)$-th stage of IFFT. Then, for each multiplied output, the remaining stages (i.e., from $(n-r+1)$-th stage to $n$-th stage) are processed and these $U$ outputs become $U$ alternative OFDM signal sequences. Among them, the OFDM signal sequence with minimum PAPR is transmitted. For selection, $\gamma$ is used similarly to the conventional SLM scheme.


\subsubsection{Wang's SLM Scheme \cite{Wang}}


In Wang's SLM scheme, the input symbol sequence $\mathbf{X}$ is IFFTed and its output, the original OFDM signal sequence $\mathbf{x}$, is multiplied by $U-1$ different $N \times N$ matrices which are called as conversion matrices to generate alternative OFDM signal sequences. Totally, the $U$ alternative OFDM signal sequences are generated. Among them, the OFDM signal sequence with minimum PAPR is transmitted. For selection, $\gamma$ is used similarly to the conventional SLM scheme.

\subsubsection{Baxley's SLM Scheme \cite{Baxely}}

The generation procedure of alternative OFDM signal sequences in Baxely's SLM scheme is the same as the case of the conventional SLM scheme. But, the selection strategy is different. For simplicity, let us assume that the HPA is linear up to the saturation point $\gamma_0$. Then, achieving a PAPR value less than $\gamma_0$ does not help to improve the system power efficiency. That is, Baxely's SLM scheme only tests phase rotation vectors until an OFDM signal sequence with PAPR less than $\gamma_0$ is found.

With overwhelming low probability, all $U$ alternative OFDM signal sequences have larger PAPR values than $\gamma_0$. In this case, Baxely's SLM scheme selects the one with minimum PAPR among them although the PAPR of that is larger than $\gamma_0$.






\section{Adaptive Generation Method of OFDM Signals in the Conventional SLM Scheme}

In this section, we introduce the AG method and its application to the conventional SLM scheme in \cite{Bauml} which is the most basic SLM scheme.

\subsection{Adaptive Generation Method of OFDM Signals}

In Pseudo code 1, when the conventional SLM scheme generates $u$-th alternative OFDM signal sequence, their components' powers and PAPR value are observed after the $u$-th alternative OFDM signal sequence, $\mathbf{x}^u=\{x^u(0),x^u(1),...,x^u(N-1)\}$, is fully generated. However, this methodology is inefficient in the computational sense.

The alternative OFDM signal sequences are successively generated from $1$-st to $U$-th as Pseudo code 1. Suppose that $\gamma$ is the intermediate minimum PAPR value up to the $(u-1)$-th alternative OFDM signal sequence. And next, while generating the $u$-th alternative OFDM signal sequence, if an OFDM signal having larger power than $\gamma E[|x(n)|^2]$ is generated, we can stop this generation procedure immediately. The reason is that, in this case, this $u$-th alternative OFDM signal sequence is never selected. Of course, the PAPR reduction performance of the conventional SLM scheme is not effected by this stop.

\begin{figure}[H]
\centering
\includegraphics[width=0.9\linewidth]{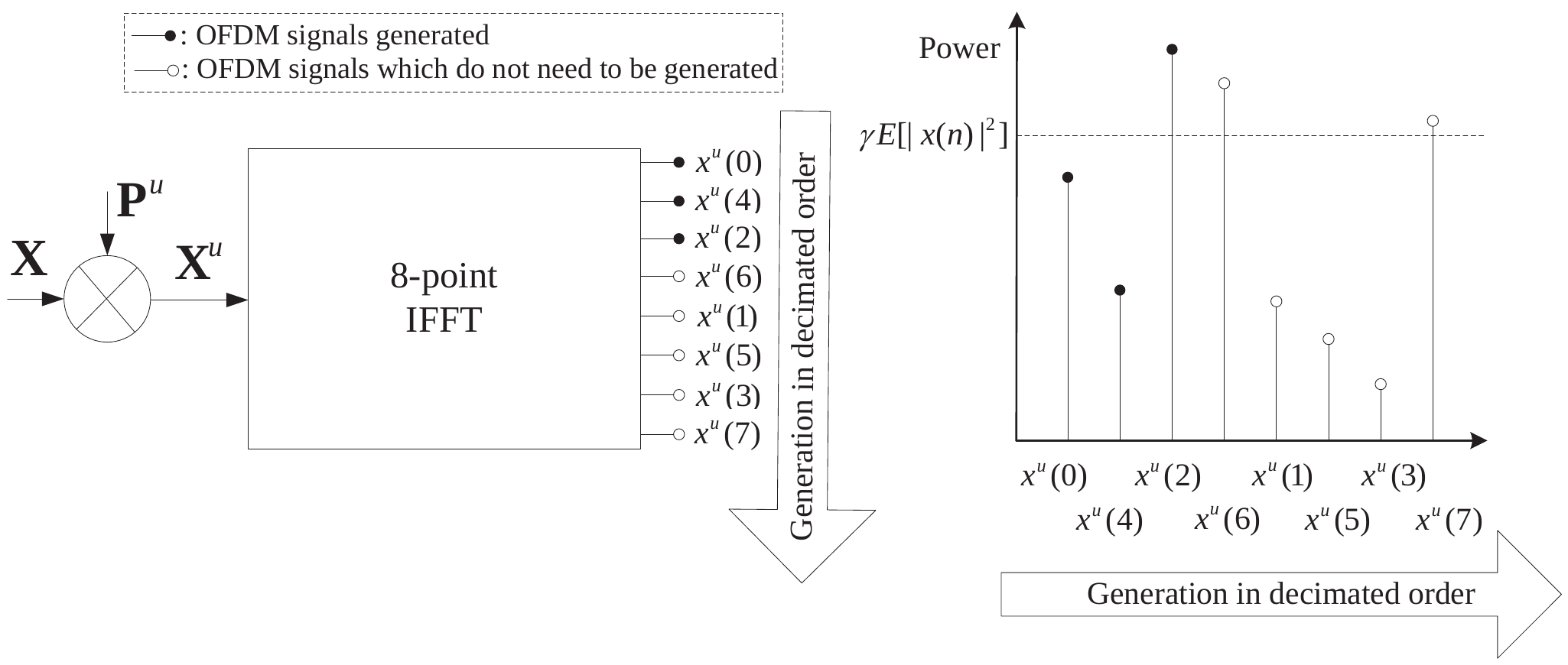}
\caption{An example of the AG method at an IFFT in the conventional SLM scheme when $N=8$.}
\label{fig:AGconcept}
\end{figure}

Fig. \ref{fig:AGconcept} represents this AG method pictorially. In Fig. \ref{fig:AGconcept}, the $u$-th decimation-in-time (DIT) IFFT block in Fig. \ref{fig:convSLM} is shown when $N=8$. The IFFT block generates the $u$-th alternative OFDM signals in decimated index order. And the generating procedure can be stopped immediately after the third OFDM signal, $x^u(2)$, is generated because $x^u(2)$ has larger power than $\gamma E[|x(n)|^2]$. Without the AG method, the remaining OFDM signals $x^u(6),x^u(1),x^u(5),x^u(3),x^u(7)$ would be fully generated.

The important question is: Is it possible that a part of an alternative OFDM signal sequence can be generated by processing a part of one IFFT, that is, by lower computational complexity than the computational complexity of one IFFT. If possible, ``stop'' in the AG method can remove the unnecessary computational complexity of the conventional SLM scheme. The answer is ``possible''. And, similarly, the cases of almost all SLM schemes including Lim's, Wang's, and Baxley's SLM schemes are possible, too.

\subsection{Partial Generation of OFDM Signals by Partially Processing an IFFT}

\begin{figure}[H]
\centering
\includegraphics[width=0.9\linewidth]{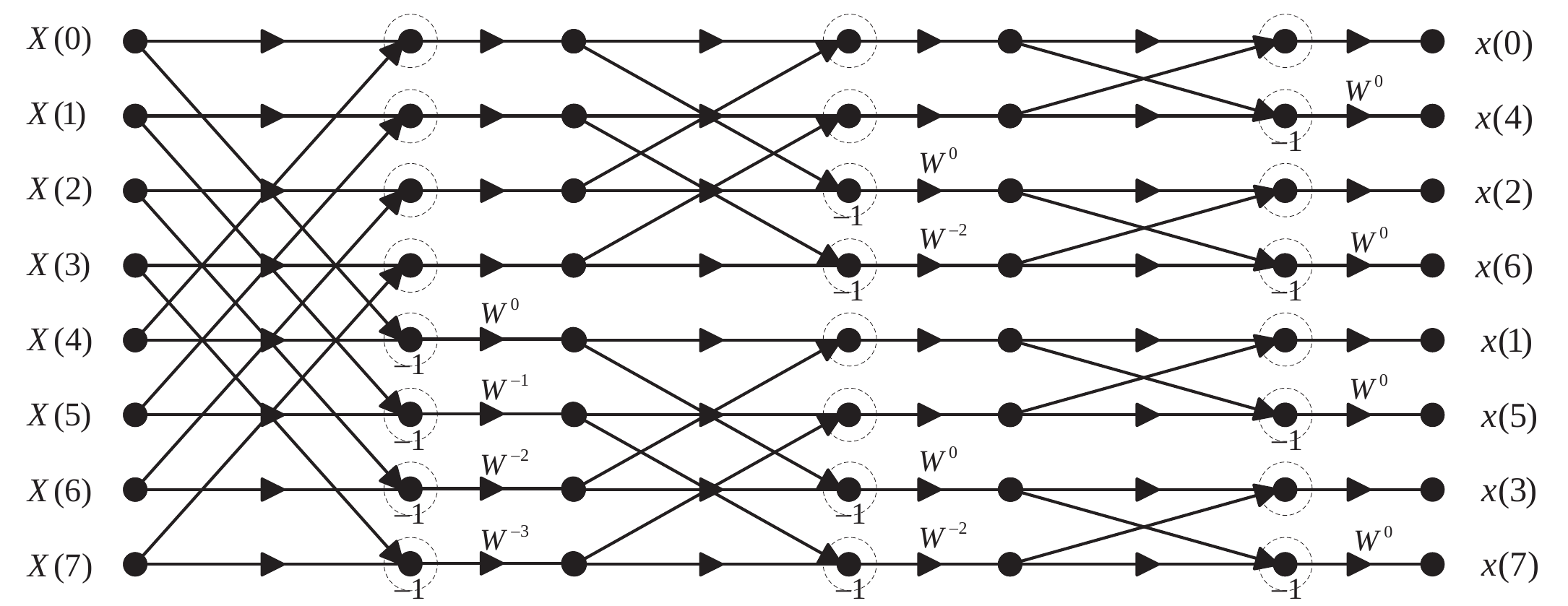}
\caption{An $8$-point IFFT structure in DIT and its c-points.}
\label{fig:DIT_IFFT}
\end{figure}

In the computational sense, it is known that the $N$-point IFFT has totally $N \log_2 N$ points which have to be computed by some complex additions and/or multiplications. Generally, the computational complexity of the IFFT is induced from these points and we call them as c-points in this paper. Fig. \ref{fig:DIT_IFFT} shows a $8$-point IFFT structure in DIT and there are $24$ c-points marked by dashed circles.

\begin{figure}[H]
\centering
\includegraphics[width=0.9\linewidth]{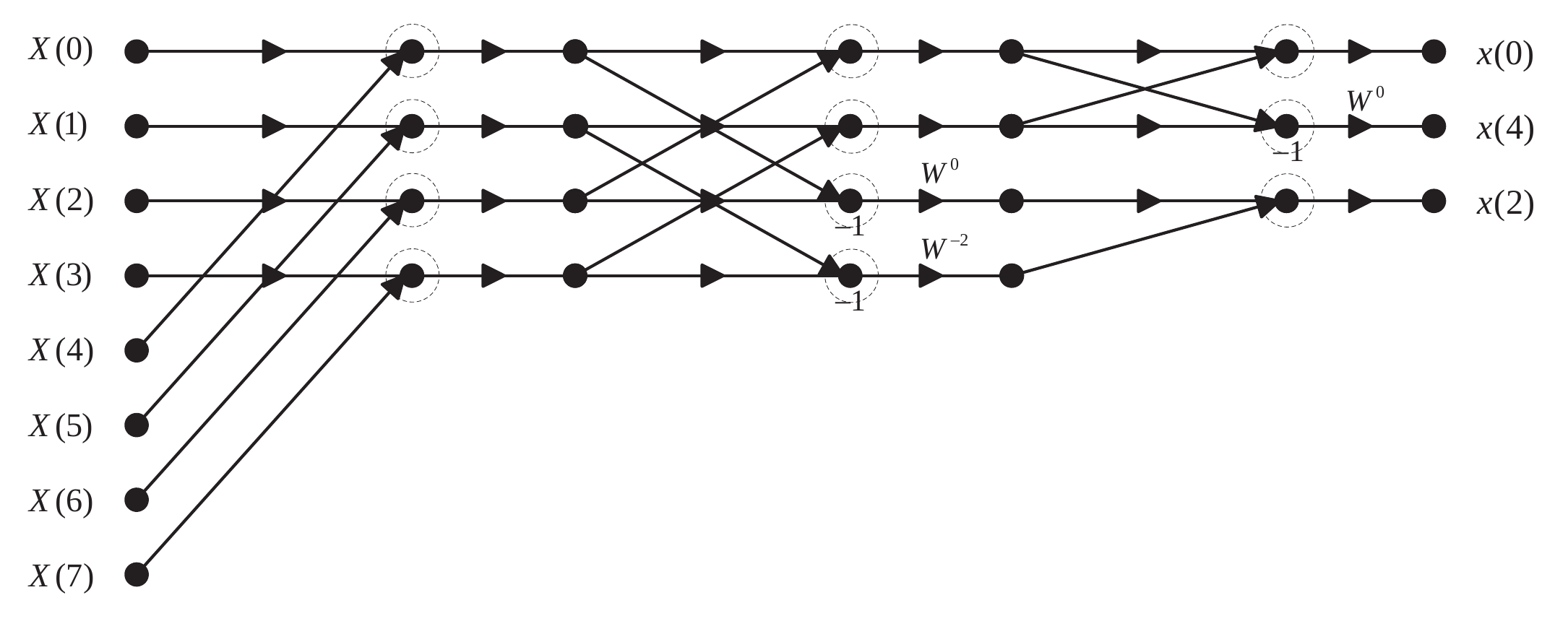}
\caption{The required part of the $8$-point IFFT to generate the OFDM signals $x(0), x(4), x(2)$.}
\label{fig:partialFFT}
\end{figure}

In the structure of the IFFT, some part of the OFDM signals can be generated by processing the corresponding part of the IFFT. For instance, in the Fig. \ref{fig:partialFFT}, $x(0),x(4),x(2)$ can be generated by not the full IFFT but the partial IFFT. That is, $11$ c-points can be computed instead of $24$ c-points to generate these three OFDM signals.

In analogous ways, we can also successively generate the OFDM signals $x(0),x(4),x(2)$ by one and one as follows. Firstly, $x(0)$ is generated by computing the seven c-points. Next, $x(4)$ is generated by computing the one c=point additively to the c-points for $x(0)$. Finally, $x(2)$ is generated by computing the three c-points additively to the c-points for $x(4)$.

\subsection{Conventional SLM Scheme with the Proposed AG Method}

By combining the contents of subsections III-A and III-B, we describe the conventional SLM scheme with the AG method.
Pseudo code 2 shows the detailed procedure of the conventional SLM scheme aided by the AG method. At third and fourth lines in Pseudo code 2, the alternative OFDM signals are generated in decimated index order. That is, the OFDM signals of the $u$-th OFDM signal sequence are successively generated by one and one by computing the required part of the IFFT additively as we described in subsection III-B. As fifth line in Pseudo code 2, the generation procedure can be stopped based on the value of $\gamma$. Then, we can remove the unnecessary  computational complexity from fully generating the alternative OFDM signal sequences.
\begin{flushleft}
\rule{.99\linewidth}{0.2mm}\\
\end{flushleft}
\textbf{Pseudo code 2: the conventional SLM scheme aided by the AG method}\\
1:~~$\gamma \Leftarrow \infty$.\\
2:~~\textbf{for} $u=1, 2, \cdots, U$\\
3:~~~~~~~\textbf{for} $n=0, N/2, N/4, \cdots, N-1$\\
4:~~~~~~~~~~~~generate $x^u(n)$ by processing the required part of one $N$-point IFFT additively.\\
5:~~~~~~~~~~~~\textbf{if} $|x^u(n)|^2/E[|x(n)|^2] > \gamma$\\
6:~~~~~~~~~~~~~~~~~go to 11.\\
7:~~~~~~~~~~~~\textbf{end if}\\
8:~~~~~~~\textbf{end for}(n)\\
9:~~~~~~~$\gamma \Leftarrow$ PAPR of $\mathbf{x}^u$.\\
10:~~~~~~$\mathbf{x}^{\tilde{u}} \Leftarrow$ $\mathbf{x}^u$.\\
11:~\textbf{end for}(u)\\
12:~transmit $\mathbf{x}^{\tilde{u}}$.\\
\rule{.99\linewidth}{0.2mm}\\

\begin{figure}[H]
\centering
\includegraphics[width=.9\linewidth]{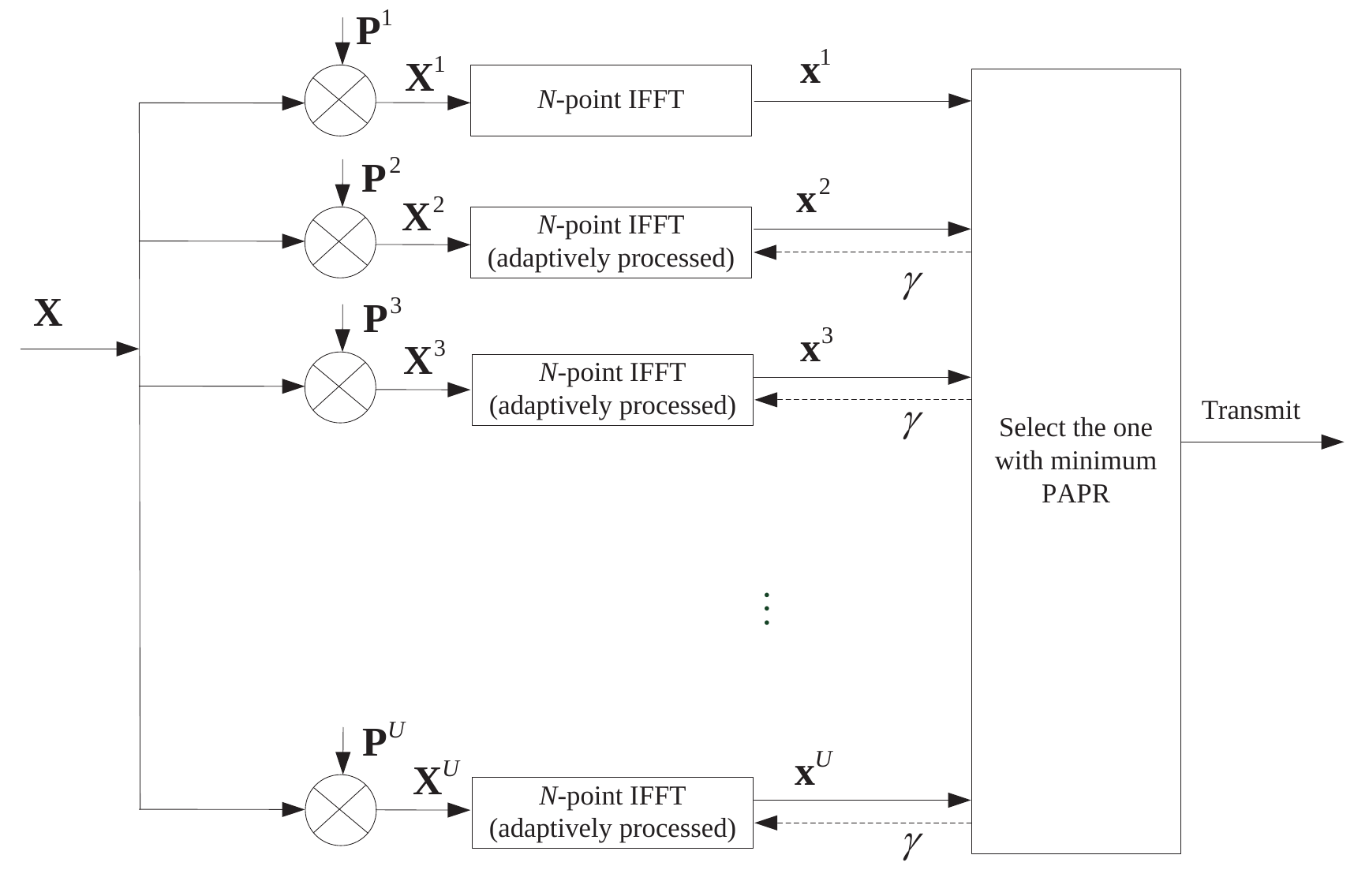}
\caption{A block diagram of the conventional SLM scheme aided by the proposed AG method.}
\label{fig:AGconvSLM}
\end{figure}

Fig. \ref{fig:AGconvSLM} shows a block diagram of the conventional SLM scheme aided by the AG method. Except the first IFFT block, at the each IFFT block, generation of the each alternative OFDM signal sequence is adaptively processed based on the value of $\gamma$ and can be stopped during the generation procedure.

\subsection{Computational Complexity Analysis}

In this subsection, we analyze the computational complexity of the conventional SLM scheme with the AG method stochastically. As other literatures, we only consider the computational complexity to generate alternative OFDM signal sequences.

\subsubsection{Computational Complexity for Partially Processing an IFFT}
In this subsubsection, we describe the computational complexity required to process a partial IFFT. We regard the computational complexity required to process fully one $N$-point IFFT as $T$. And there are $N \log_2N$ c-points have to be computed in one $N$-point IFFT. We define roughly the computational complexity for one c-point as
\begin{equation}
t \triangleq \frac{T}{N \log_2 N}.
\end{equation}

And, considering the $N$-point IFFT structure consisting of smaller-point IFFTs, the computational complexity to generate $a, 1 \leq a \leq N$, OFDM signals in the way of subsection III-B is easily obtained by
\begin{equation}\label{eq:K(a)}
K(a) = \{2^0+\cdots+2^{n-1}\}t + \{ \lfloor\frac{a-1}{2^0}\rfloor\cdot2^0 + \cdots + \lfloor\frac{a-1}{2^{n-1}}\rfloor\cdot2^{n-1}  \}t
\end{equation}
where $n=log_2N$. Trivially, $K(N) = N \log_2 N t = T$ which means processing fully one $N$-point IFFT.

\begin{figure}[H]
\centering
\includegraphics[width=0.9\linewidth]{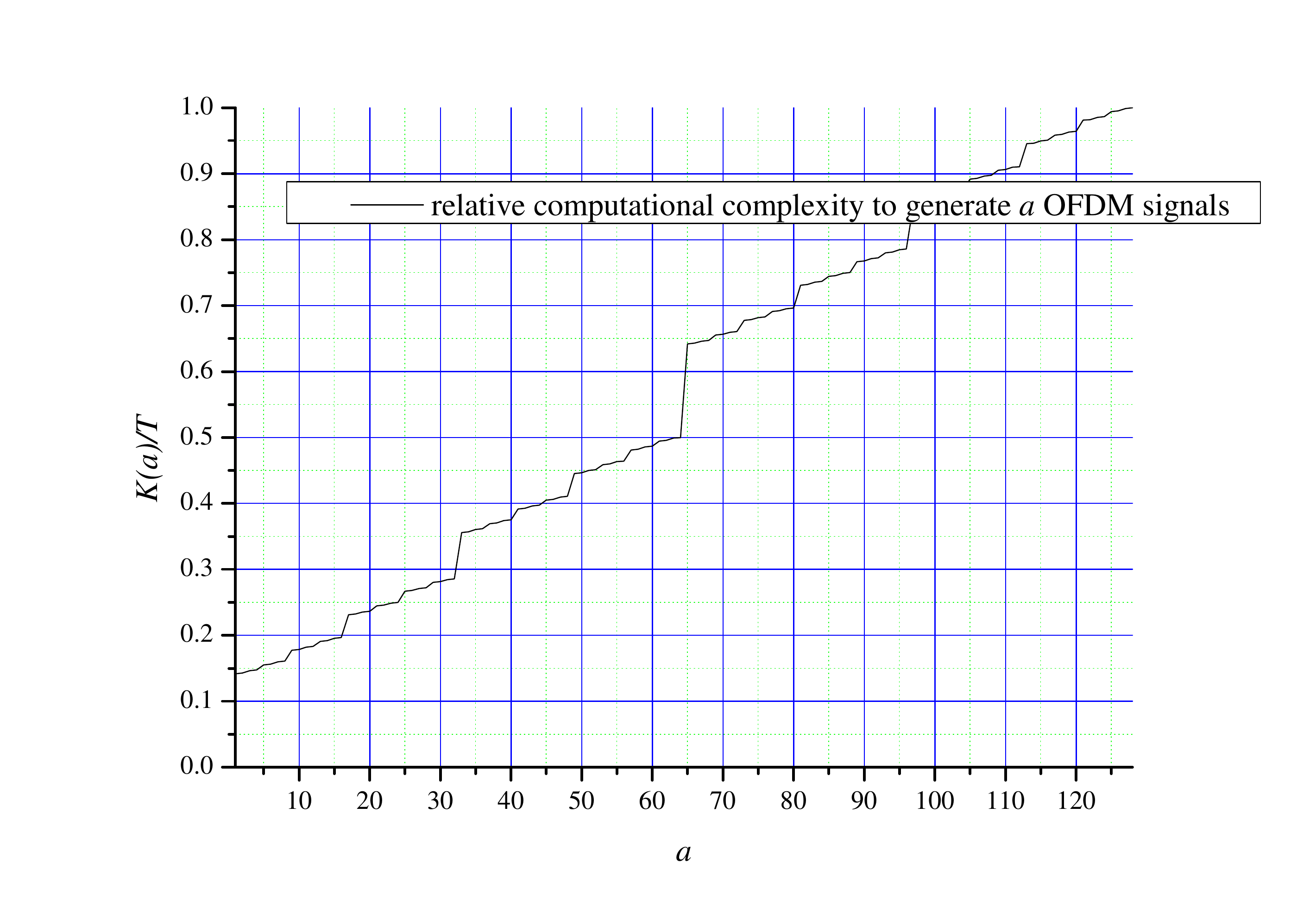}
\caption{Relative computational complexity required to generate $a$ OFDM signals at a $128$-point IFFT.}
\label{fig:plotAG}
\end{figure}
To have insight in the computational sense, $K(a)/T$ versus $a$ is plotted as Fig. \ref{fig:plotAG} for a $128$-point IFFT. It is remarkable that the plot in Fig. \ref{fig:plotAG} behaves in a near linear fashion. For instance, $a=64$ at the $x$-axis corresponds to $0.5$ at the $y$-axis, which means that one can generate a half of an OFDM signal sequence by a half cost of the computational complexity to process one $128$-point IFFT.

\subsubsection{Probability Distribution of the Number of Alternative OFDM Signals Generated}

In this subsubsection, we derive the probability distribution of the number of alternative OFDM signals generated at the each IFFT block in the conventional SLM scheme when the AG method is applied.

We remark the fundamental issues. In this paper, our analysis is based on the heuristic assumption that the baseband OFDM signal is characterized as a complex Gaussian, which becomes accurate as the number of subcarriers increases due to the central limit theorem \cite{Ochiai}. Now, assuming that the OFDM signal is complex Gaussian, the amplitude of the OFDM signal is Rayleigh distributed and it can be easily shown that the probability that one OFDM signal $x(n)$ is smaller than $\gamma E[|x(n)|^2]$ is denoted as $\Gamma$ and given by
\begin{align}
\Gamma \triangleq & P\bigg(\frac{|x(n)|^2}{E[|x(n)|^2]} < \gamma \bigg)\nonumber\\
 =& 1-e^{-\gamma}.
\end{align}

Moreover, as the assumption in \cite{Bauml}, we assume that the components in an OFDM signal sequence are mutually independent from the central limit theorem. And we also assume that $U$ alternative OFDM signal sequences are mutually independent. The above assumptions are valid only when the Nyquist sampling rate is used in OFDM systems.

We denote the number of OFDM signals generated at the $u$-th IFFT block as a random variable $A_u$ for $2 \leq u\leq U$. The distribution of $A_u$ depends on the value of $\gamma$ after $(u-1)$-th alternative OFDM signal sequence is generated. And $\gamma$ is equal to the PAPR value of the conventional SLM scheme with $u-1$ alternative OFDM signal sequences. Therefore, the probability mass functions (PMFs) of $A_u$ can be represented as
\begin{equation}\label{eq:PMFofAu}
p_{A_u}(a_u) = \int_1^{\infty} p_{A_u | PAPR_{SLM(u-1)}} (a_u | \gamma)~f_{PAPR_{SLM(u-1)}}(\gamma) d\gamma
\end{equation}
where $a_u$ is an integer within $[1,N]$, $p_{A_u | PAPR_{SLM(u-1)}} (a_u | \gamma)$ is the conditional PMF of $A_u$ given the value of $PAPR_{SLM(u-1)}$, and $f_{PAPR_{SLM(u-1)}}(\gamma)$ is the probability density function (PDF) of the random variable $PAPR_{SLM(u-1)}$. The random variable $PAPR_{SLM(u-1)}$ means the PAPR value from the conventional SLM scheme generating $u-1$ alternative OFDM signal sequences. And the range of integral in (\ref{eq:PMFofAu}) comes from the definition of PAPR.

With $\Gamma$, we describe the functions $p_{A_u | PAPR_{SLM(u-1)}} (a_u | \gamma)$ and $f_{PAPR_{SLM(u-1)}}(\gamma)$ as
\begin{equation}\label{eq:subpmf}
p_{A_u | PAPR_{SLM(u-1)}} (a_u | \gamma) = \begin{cases} \Gamma^{a_u-1}(1-\Gamma), &1\leq a_u \leq N-1\\
                                                         \Gamma^{N-1},           &a_u = N
                                           \end{cases}
\end{equation}
and
\begin{align}\label{eq:subpdf}
f_{PAPR_{SLM(u-1)}}(\gamma) &= \frac{d}{d \gamma} F_{PAPR_{SLM(u-1)}}(\gamma)\nonumber\\
                            &= \frac{d}{d \gamma} (1-(1-\Gamma^N)^{u-1}),
\end{align}
respectively, where $F_{PAPR_{SLM(u-1)}}(\gamma)$ is the cumulative distribution function (CDF) of $PAPR_{SLM(u-1)}$.

Combining (\ref{eq:K(a)}) and (\ref{eq:PMFofAu}), the expectation of the computational complexity of the conventional SLM scheme aided by the AG method is given by
\begin{equation}\label{eq:CCforconvSLMwithAG}
T+\sum_{a_2=1}^{N} K(a_2)p_{A_2}(a_2)+\cdots+\sum_{a_U=1}^{N} K(a_U)p_{A_U}(a_U).
\end{equation}
Trivially, the computational complexity of the conventional SLM scheme generating $U$ alternative OFDM signal sequences without the AG method is $UT$.

\subsubsection{Comparison between the Analytical and Numerical Results}

In this subsubsection, we compare the analytical results and the numerical results of the computational benefit of the AG method. The number of subcarriers is $N=64$ and the Nyquist sampling rate is used. And the numbers of the alternative OFDM signal sequences are $U=2,3,\cdots,16$.

\begin{figure}[H]
\centering
\includegraphics[width=0.9\linewidth]{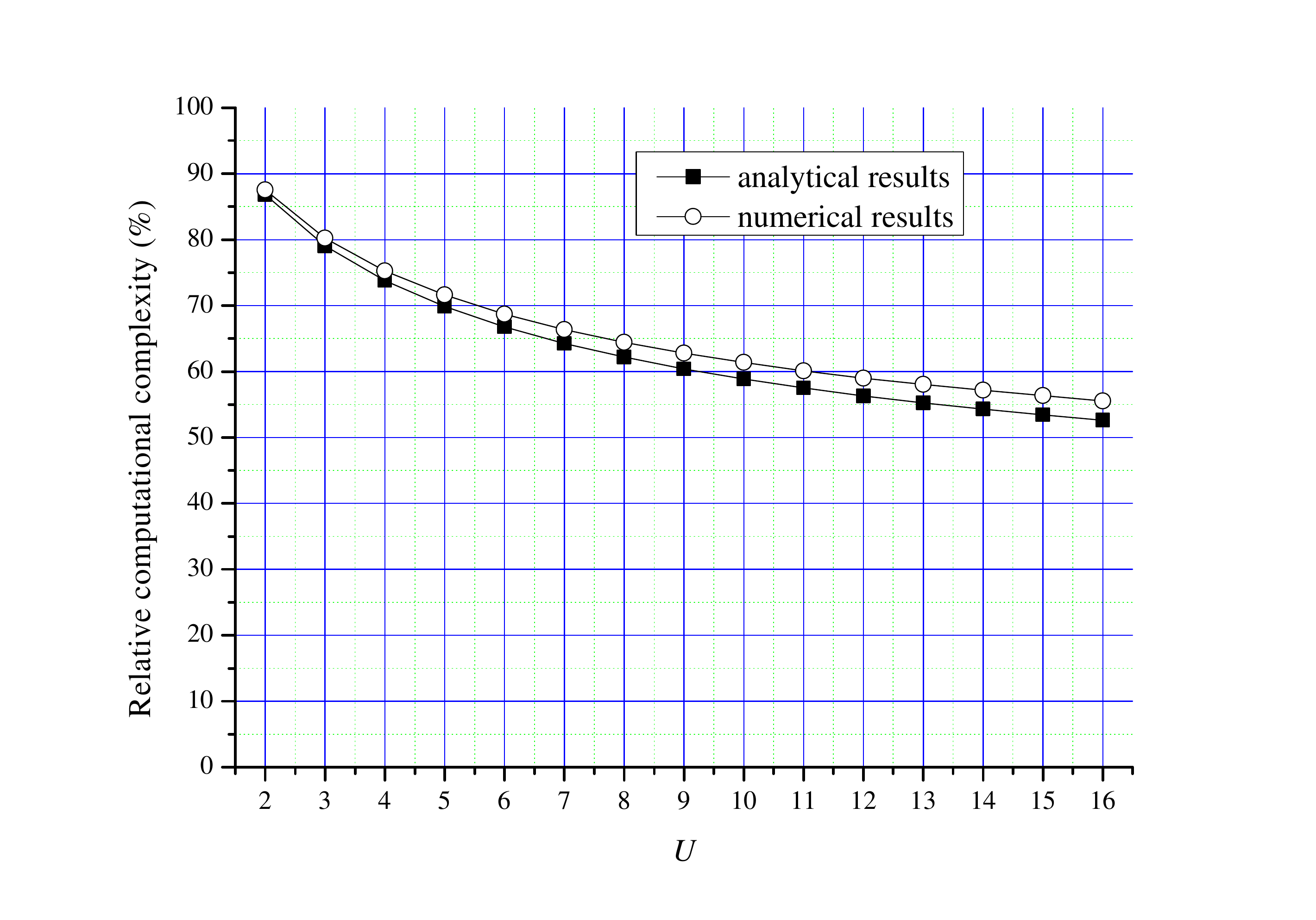}
\caption{Relative computational complexity of the conventional SLM scheme with the AG method compared to the case without the AG method for $N=64$ and various $U$.}
\label{fig:avsn}
\end{figure}
Fig. \ref{fig:avsn} shows the relative computational complexity required to the conventional SLM scheme with the AG method compared to the case without the AG method (i.e., $UT$). The analytical results are given by the value of (\ref{eq:CCforconvSLMwithAG}) to $UT$ ratio and the numerical results are given by testing $10^5$ randomly generated input symbol sequences by computer simulations. For small $U$, two results are similar. However, for large $U$, there is a little gap between the two results. This difference comes from the assumption, the $U$ alternative OFDM signal sequences are mutually independent. In practical situations, there are correlations between the $U$ alternative OFDM signal sequences.

\section{Adaptive Generation Method of OFDM Signals in the Low-complexity SLM Schemes}

In this section, we briefly introduce Lim's \cite{Lim}, Wang's \cite{Wang}, and Baxley's \cite{Baxely} SLM schemes applied by the AG method. The basic methodology of these applications are similar to the conventional SLM scheme case. The AG method can be applied to other SLM scheme beyond these SLM schemes analogously.
\subsection{Lim's SLM Scheme with the AG Method}

In Lim's SLM scheme with the AG method, the common IFFT processed from $1$-st stage to $(n-r)$-th stage is the same. But, the remaining stages of $U$ IFFTs can be adaptively processed based on the value of $\gamma$. Clearly, it is possible that a part of an alternative OFDM signal sequence can be generated by partially processing the remaining stages of the IFFT.




\subsection{Wang's SLM Scheme with the AG Method}
Clearly, some part of an alternative OFDM signal sequence can be generated by multiplying the corresponding partial columns of the conversion matrix to the original OFDM signal sequence $\mathbf{x}$. That is, the $U-1$ conversion matrix-vector multiplications can be adaptively processed based on the value of $\gamma$.




\subsection{Baxely's SLM Scheme with the AG Method}
In Baxely's SLM scheme, there is the value of $\gamma_0$, the saturation point of HPA, instead of $\gamma$. With the AG method, each IFFT is adaptively processed based on $\gamma_0$. That is, while alternative OFDM signals are being generated, the IFFT process can be stopped if an OFDM signal having PAPR larger than $\gamma_0 E[|x(n)|^2]$ is generated. Then, the next alternative OFDM signal sequence has to be generated and tested by $\gamma_0$ in the same manner.

As we described in subsection II-B, with overwhelming low probability, all the $U$ alternative OFDM signal sequences have larger PAPR values than $\gamma_0$. In this case, Baxely's SLM scheme with the AG method select the one with minimum PAPR among them by finishing the $U$ partially processed IFFT blocks.




\section{Numerical Results}

In this section, we present some numerical results. The OFDM signal sequences are four times oversampled by inserting zeroes into (alternative) input symbol sequences. We obtain the numerical results from the simulations to present the computational benefit of the proposed AG method exactly. The results are given by testing $10^5$ randomly generated input symbol sequences. As other literatures, the computational complexity to generate alternative OFDM signal sequences is only considered for the comparison. Clearly, the PAPR reduction performance is not degraded by the AG method and thus we compare only the computational complexity. The reduction of the computational complexity does not depend on the modulation order and $16$-quadrature amplitude modulation ($16$-QAM) is used for all simulations.\\

\subsection{Numerical Results for the Conventional SLM Scheme}

For the conventional SLM scheme, we simulate the OFDM system when $N=256$ and $N=1024$. Table I shows the computational complexity of the conventional SLM scheme with the AG method and that without the AG method. As we mentioned, the computational complexity of the conventional SLM scheme without the AG method is $UT$. Table I shows that the AG method can reduce the computational complexity of the conventional SLM scheme substantially.

For instance, when $U=32$, the computational complexity of the conventional SLM scheme aided by the AG method is reduced to be almost $33\%$ of its original cost. Table I shows that the number of subcarriers $N$ does not effect the computational benefit of the AG method. And the AG method has a large benefit as $U$ increases.
\begin{table}[!h]
\centering
\caption{The Computational Benefit of the AG Method for the Conventional SLM Scheme.}
\begin{tabular}{|c|c|c|c|c|}
\hline
\multicolumn{2}{|c|}{}& $U=8$ & $U=16$ & $U=32$\\
\hline
        & Conventional SLM without AG (a) & $8T$ & $16T$ & $32T$\\
\cline{2-5}
$N=256$ & Conventional SLM with AG (b) & $4.21T$ & $6.69T$ & $10.82T$\\
\cline{2-5}
        & (b)/(a) $(\%)$ & 52.6 & 41.8 & 33.8\\
\hline
        & Conventional SLM without AG (c) & $8T$ & $16T$ & $32T$ \\
\cline{2-5}
$N=1024$& Conventional SLM with AG (d) & $4.22T$ & $6.65T$ & $10.70T$\\
\cline{2-5}
        & (d)/(c) $(\%)$ & 52.7 & 41.6 & 33.4\\
\hline
\end{tabular}
\end{table}
\\

\subsection{Numerical Results for the Low-complexity SLM Schemes}

Beyond the conventional SLM scheme, we also present the numerical results when we apply the AG method to the three low-complexity SLM schemes. The number of subcarriers is fixed to $N=256$.

Table II shows the computational complexity of Lim's SLM scheme with the AG method and that without the AG method. In Lim's SLM scheme, the number of the remaining stages is $r=5$ which is the guaranteed value to have good PAPR reduction performance. The AG method can reduce the computational complexity substantially of Lim's SLM scheme.
\begin{table}[!h]
\centering
\caption{The Computational Benefit of the AG Method for Lim's SLM Scheme \cite{Lim}.}
\begin{tabular}{|c|c|c|c|}
\hline
           &     $U=8$ &    $U=16$ &    $U=32$ \\
\hline
Lim's SLM without AG (a) &  $4.5T$    &  $8.5T$   &   $16.5T$  \\
\hline
Lim's SLM with AG (b) & $2.46T$    & $3.48T$    & $5.10T$    \\
\hline
(b)/(a) ($\%$) & 54.7  & 40.9  &  30.9  \\
\hline
\end{tabular}
\end{table}

Table III shows the computational benefit when the AG method is applied to Wang's SLM scheme. In Table III, we present the comparison of the number of complex additions which are required for conversion matrix-vector multiplications in Wang's SLM scheme. And Wang's SLM scheme has a constraint on $U$ and the cases of $U=4,8,12$ are simulated.
\begin{table}[!h]
\centering
\caption{The Computational Benefit of the AG Method for Wang's SLM Scheme \cite{Wang}.}
\begin{tabular}{|c|c|c|c|}
\hline
           &     $U=4$ &    $U=8$ &    $U=12$ \\
\hline
Wang's SLM without AG (a) &  9,216    &  21,504   & 33,792  \\
\hline
Wang's SLM with AG (b) & 4,933    & 9,288    & 12,820    \\
\hline
(b)/(a) ($\%$) & 53.5  & 43.2  &  37.9  \\
\hline
\end{tabular}
\end{table}

Table IV shows the computational complexity when the AG method is applied to Baxely's SLM scheme. In this case, the number of alternative OFDM signal sequences is fixed to $U=16$. In Table IV, the computational benefit of the AG method depends on the value of $\gamma_0$, the saturation point of HPA. For $\gamma_0=8.0$dB, the proposed AG method can reduce the computational complexity of Baxely's SLM scheme to be almost $55\%$ of its original cost.
\begin{table}[!h]
\centering
\caption{The Computational Benefit of the AG Method for Baxely's SLM Scheme \cite{Baxely}.}
\begin{tabular}{|c|c|c|c|}
\hline
                                   &  $\gamma_0=7.5$dB & $\gamma_0=8.0$dB & $\gamma_0=8.5$dB \\
\hline
Baxely's SLM scheme without AG (a) & $8.03T$ & $3.24T$ & $1.73T$ \\
\hline
Baxely's SLM scheme with AG (b)    & $5.12T$ & $1.81T$ & $1.28T$ \\
\hline
(b)/(a) ($\%$)                     & $63.8$ & $55.9$ & $73.9$ \\
\hline
\end{tabular}
\end{table}

It is remarkable that the AG method has a large computational benefit when we apply the AG method to the low-complexity SLM schemes which are already modified for low-complexity. That is, the AG method can be combined effectively to almost all existing SLM schemes to have much lower computational complexity.

\section{Conclusions}
There are many SLM schemes including the conventional SLM scheme and its low-complexity versions. The SLM schemes generate the alternative OFDM signal sequences. We proposed the AG method of the alternative OFDM signals in the SLM schemes. With the AG method, the computational complexity of the SLM schemes can be reduced substantially. It is meaningful that the application of the AG method does not degrade any PAPR reduction performance of the SLM scheme. In this paper, we described the application for the conventional SLM scheme and its stochastic analysis is also given. And, we briefly described the applications for the three representative low-complexity SLM schemes. Numerical results show the computational benefit of using the AG method. It is remarkable that the AG method is also effective for the low-complexity SLM scheme which is already modified to have low-complexity. We anticipate that the AG method can be applied to many other SLM schemes beyond the SLM schemes described in this paper.

\section*{Acknowledgment}
This work was supported by the National Research Foundation of Korea (NRF) grant funded by the Korea government (MEST) (No. 2012-0000186).

\end{document}